\documentclass[twocolumn,times,tight]{aastex62}
\graphicspath{{./}{figures/}}
\usepackage{CJK}

\usepackage{amsmath}
\usepackage{xcolor}

\shorttitle{Electric Current Neutralization in Solar Active Regions}
\shortauthors{Avallone et al.}

\begin{document}

\begin{CJK*}{UTF8}{gbsn}

\title{Electric Current Neutralization in Solar Active Regions and Its Relation to Eruptive Activity}

\author[0000-0003-1719-5046]{Ellis A. Avallone}
\affil{Institute for Astronomy, University of Hawai'i at M\={a}noa, 2680 Woodlawn Dr., Honolulu, HI 96822, USA}

\author[0000-0003-4043-616X]{Xudong Sun (孙旭东)}
\affil{Institute for Astronomy, University of Hawai`i at M\={a}noa, Pukalani, HI 96768, USA}

\correspondingauthor{Xudong Sun (孙旭东)}
\email{xudongs@hawaii.edu}

\begin{abstract}
     It is well established that magnetic free energy associated with electric currents powers solar flares and coronal mass ejections (CMEs) from solar active regions (ARs). However, the conditions that determine whether an AR will produce an eruption are not well understood. Previous work suggests that the degree to which the driving electric currents, or the sum of all currents within a single magnetic polarity, are neutralized may serve as a good proxy for assessing the ability of ARs to produce eruptions. Here, we investigate the relationship between current neutralization and flare/CME production using a sample of 15 flare-active and 15 flare-quiet ARs. All flare-quiet and 4 flare-active ARs are also CME-quiet. We additionally test the relation of current neutralization to the degree of shear along polarity inversion lines (PILs) in an AR. We find that flare-productive ARs are more likely to exhibit non-neutralized currents, specifically those that also produce a CME. We find that flare/CME-active ARs also exhibit higher degrees of PIL shear than flare/CME-quiet ARs. We additionally observe that currents become more neutralized during magnetic flux emergence in flare-quiet ARs. Our investigation suggests that current neutralization in ARs is indicative of their eruptive potential.
\end{abstract}

\section{Introduction}\label{intro}
Solar active regions (ARs) harbor strong magnetic fields and often host sunspots. They are a major source of eruptive activity, including solar flares and coronal mass ejections (CMEs). Solar eruptions are drivers of space weather events \textemdash\, changes in near-Earth space that can negatively affect the technology we use on Earth (i.e., telecommunications satellites). A substantial effort has been made to accurately predict space weather events \citep[see e.g.,][]{2015ApJ...798..135B,2007ApJ...656.1173L}. Such predictions are heavily dependent on understanding the source environments of solar eruptions (i.e., ARs).

\begin{figure*}[ht!]
    \centering
    \includegraphics[width=0.9\textwidth]{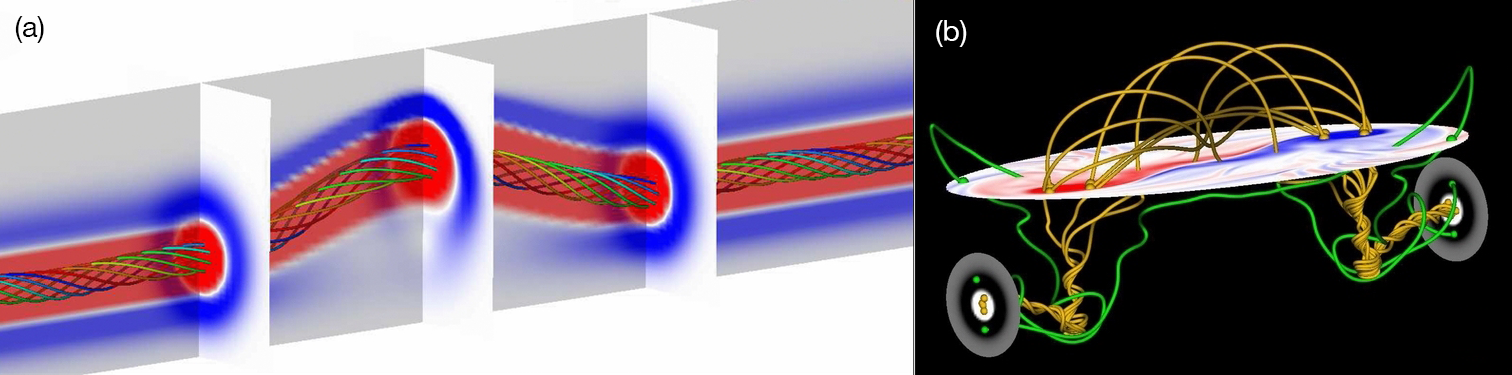}
    \caption{Figure from \citet{2014ApJ...782L..10T} showing a bipolar AR comprised of a flux rope. Panel (a) shows a cartoon of a pre-emergence flux rope, its associated magnetic field lines (rainbow), and direct (orange) and return (blue) currents. Panel (b) shows the results from a numerical simulation of AR formation from \citet{2013ApJ...778...99L} and highlights the distribution of electric current streamlines post-flux emergence in an AR. The direct currents are shown in orange and the return currents are shown in green. The return currents are trapped beneath the photosphere after magnetic flux emerges.}
    \label{flux_rope}
\end{figure*}

The standard model for an AR states that it is comprised of a flux rope \textemdash\, a tube-like region of space containing a twisted magnetic field. An example of a flux rope is shown by the cartoon in Figure~\ref{flux_rope}(a). When a twisted flux tube emerges through the photosphere, an AR will form, indicated by a strong concentration of magnetic field in the photosphere. Flux ropes are also current carrying structures. Direct currents (DC) connect the centers of each AR polarity and are generated by the flux rope itself, while return currents (RC) surround the direct currents and oppose them. Figure~\ref{flux_rope}(a) shows where these currents are nominally distributed in a flux rope prior to emergence through the photosphere.

It is well-established that free magnetic energy associated with electric currents in ARs drives flare/CME eruptions \citep[see e.g.,][]{2005ApJ...628..501S,2011LRSP....8....6S}. However, it is still unclear how well these driving electric currents are \textit{neutralized}. An AR is considered neutralized when the ratio of DC to RC \(|DC/RC|\) in each magnetic polarity is close to 1.

The degree of current neutralization in ARs has been debated for several decades. \citet{1996ApJ...471..485P} suggested that the average net current density of an AR is zero. This follows from the assumption that AR magnetic fields are composed of individually current-neutralized magnetic fibrils that, when combined, render the entire magnetic configuration current-neutralized. However, \citet{1991ApJ...381..306M,1995ApJ...451..391M} argued that non-neutralized currents can emerge along with the emergence of magnetic flux as an AR evolves. \citet{1991ApJ...381..306M,1995ApJ...451..391M} also suggested that the return currents could escape detection, which would render the net currents non-neutral. \citet{2000ApJ...545.1089L} devised a model for this by showing that return currents become trapped beneath the photosphere as magnetic flux emerges, thus accounting for hidden return currents, as illustrated in the numerical simulation output in Figure~\ref{flux_rope}(b). \citet{2000ApJ...540.1150W} found that most of the simulated ARs in their study were non-neutralized, further supporting Melrose's scenario. More recent numerical magnetohydrodynamic (MHD) simulations support Melrose's scenario as well, showing that non-neutralized currents emerge alongside the development of substantial magnetic shear along an AR's polarity inversion line (PIL), the line that divides dominant magnetic polarities in ARs \citep[e.g.,][]{2012ApJ...761...61G,2014ApJ...782L..10T,2015ApJ...810...17D}. These studies also suggest a link between PIL shear \textemdash\, the average angle between the observed vector magnetic field and a model force-free potential magnetic field near the PIL \textemdash\, and flare/CME productivity in ARs.

It has also been suggested that CME productivity may depend on the distribution of direct and return currents in ARs \citep{forbes_2010}. Previous CME simulations have utilized flux rope configurations which do not contain return currents \citep[e.g.,][]{2005ApJ...630L..97T,2008ApJ...684.1448M}. The presence of significant PIL shear in ARs prior to CME eruption implies the existence of non-potential magnetic fields, indicating the presence of the magnetic free energy required for eruption. The simulations mentioned in the previous paragraph also indicate that PIL shear develops along with non-neutralized currents. However, if currents are assumed to be neutralized, these requirements for CME eruption may not be suitable \citep{forbes_2010}. 

The aforementioned theoretical studies suggest a relationship between current neutralization, PIL shear, and flare/CME productivity. To understand this relationship further, we base our work on a pilot observational study by \citet{2017ApJ...846L...6L}, who used a sample of 4 ARs, 2 of which were emerging and 2 of which were well-developed. The authors found that CME-active ARs had non-neutralized currents in both individual polarities and the entire AR, while CME-quiet ARs were close to neutral (i.e., \(|DC/RC|\) was close to 1). The authors also found that the difference in PIL shear between CME-active and CME-quiet ARs was much less pronounced, suggesting that the degree of current neutralization in ARs may serve as a better proxy for determining the ability of ARs to produce CMEs than PIL shear. However, the sample used by \citet{2017ApJ...846L...6L} was small, and more work needs to be done to further understand how these parameters are related to flare/CME productivity in ARs. 

To expand on the pilot study by \citet{2017ApJ...846L...6L}, we track 15 flare-active and 15 flare-quiet ARs across the solar disk and determine the degree of current neutralization (\(|DC/RC|\)) and PIL shear over each AR's on-disk lifetime. Among the 15 flare-active ARs, 11 are also CME-active; all other ARs are CME-quiet (see Table~\ref{dat_tbl}).  We additionally conduct a statistical analysis of our measured parameters and assess both the statistical and systematic uncertainties in our results. In doing so, we place further constraints on whether electric current neutralization is a good proxy for assessing the ability of ARs to produce flares/CMEs. 

\section{Data \& Methods}\label{data_n_methods}
\subsection{Data}\label{data}
We consider a sample of 15 flare-active and 15 flare-quiet ARs. We define flare-active ARs as ARs which produce flares greater than or equal to M-class (i.e., flares with X-ray fluxes greater than $10^{-5}$~$\mathrm{W}/\mathrm{m}^2$) and define flare-quiet ARs as ARs which do not produce flares larger than C-class (i.e., flares with X-ray fluxes less than $10^{-5}$~$\mathrm{W}/\mathrm{m}^2$). To select our sample of ARs, we use the HELIO Solar Activity Archive\footnote{http://helio-vo.eu/solar\_activity/arstats-archive/}, which produces daily reports of the flares produced by a given AR. 

\begin{table}[t!]
    \centering
    \begin{tabular}{ccc}
         & Flare-Active & Flare-Quiet \\ \hline
       CME-Active & 11 & 0 \\
       CME-Quiet & 4 & 15 \\
    \end{tabular}
    \caption{Table showing the selection criteria of the 30 ARs in our sample. A small subset of the ARs in this sample represent extreme ends of these categories (i.e., producing many X-class flares or producing no flares, or producing many CMEs or no CMEs).}
    \label{dat_tbl}
\end{table}

To assess the flare productivity for each AR, we use solar X-ray flux data from the \textit{Geostationary Operational Environmental Satellite} (GOES)\footnote{https://www.swpc.noaa.gov/products/goes-x-ray-flux} to determine a 'flare index' (FI) for each AR \citep[e.g.,][]{2007ApJ...655L.117S}. The FI for each AR is derived by adding the number of flares produced by the AR weighted by the X-Ray magnitude of each flare. C-class flares have a weight of 1, M-class flares have a weight of 10, and X-class flares have a weight of 100. We also use several online CME catalogs \citep[e.g.,][]{2009ApJ...691.1222R,2008SoPh..248..485O} and Table 1 in \citet{2017ApJ...834...56T} to assess the CME-productivity of each AR. All online catalogs are open-access. The distribution of our selection criteria among ARs in our sample is shown in Table \ref{dat_tbl}.

To determine the degree of current neutralization, PIL magnetic shear, and the unsigned magnetic flux of all ARs in our sample, we use vector magnetogram data \citep{2014SoPh..289.3483H} from the \textit{Helioseismic and Magnetic Imager} \citep[HMI;][]{2012SoPh..275..207S} on the \textit{Solar Dynamics Observatory} \citep[\textit{SDO};][]{2012SoPh..275....3P}, which generates vector magnetograms every 720 seconds since 2010. We specifically use the Space Weather HMI Active Region Patch (SHARP) dataset \citep{2014SoPh..289.3549B}, which automatically tracks strong concentrations of photospheric magnetic field observed by HMI across the solar disk. SHARP provides magnetogram maps of the three spatial components of the magnetic field (\(B_{x},B_{y},B_{z}\)) and each component's associated error. We download data for our selected ARs at an hourly cadence. These data are available from the Joint Science Operations Center (JSOC)\footnote{http://jsoc.stanford.edu/}.

\begin{figure*}
    \centering
    \includegraphics[width=0.7\textwidth]{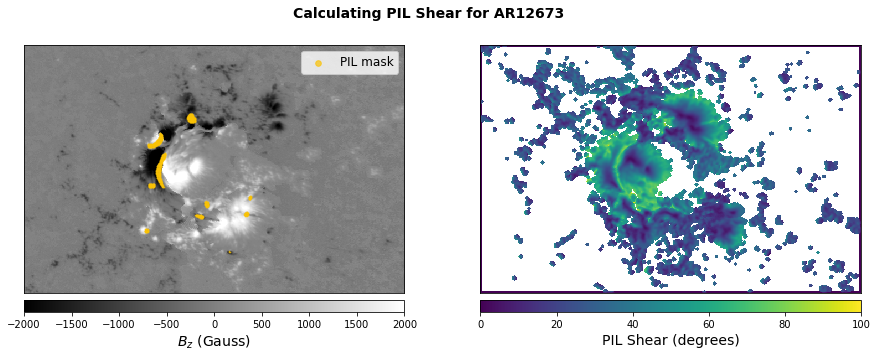}
    \caption{Computing PIL shear for flare-active AR 12673. $B_z$ magnetogram is shown on the left with its associated PIL mask (yellow). The map of shear values is shown on the right; the strongest shear is located near the PIL. Values in the weak field region ($B<200~\mathrm{G}$) are shown in white. The value for shear in Table \ref{tab:big_tbl} is computed by taking a mean of shear values within the PIL mask weighted by the magnetic field strength. Extreme shear values in regions where $B_z$ is close to zero are not included in the calculation of PIL shear.}
    \label{shear_fig}
\end{figure*}

\subsection{Calculation of Current Neutralization}\label{dcrc_calc}
To measure the degree of current neutralization (\(|DC/RC|\)) for each AR, we first derive the vertical component of the current density \(J_z\) from each HMI SHARP vector magnetogram using Amp\`ere's Law \citep{maxwell_1856} shown in Equation \ref{ampere},

\begin{equation}
J_{z} = \frac{1}{\mu _{0}} \left (\frac{\partial B_{y}}{\partial x} - \frac{\partial B_{x}}{\partial y} \right ) 
\label{ampere}
\end{equation}

\(|DC/RC|\) over each AR's on-disk lifetime is then calculated from our derived \(J_z\) maps. To do this, we first isolate the positive and negative currents within each magnetic polarity. We only consider regions where the absolute value of the magnetic field is stronger than 200~G to avoid regions with low signal-to-noise. We then calculate the net positive and net negative currents within each polarity.

To determine the sign associated with DC and DC, we determine the dominant sign of \(J_z\) in each magnetic polarity near the PIL, above which the direct current channel likely resides. There are some ARs where this association is less clear (see e.g., the right middle panel in Figure \ref{fig:pretty_ars}). However, as shown in \citet{2017ApJ...846L...6L}, these ARs are often neutralized, thus making the choice of sign for DC and RC interchangeable. Once the correct sign of DC and RC is determined, we calculate \(|DC/RC|\) in each magnetic polarity. This calculation is described further in Equations \ref{DC} and \ref{RC}. 

\begin{equation}
|DC|^{+,-} = \left\{\begin{matrix}
\left |\oint_{B_z^{+,-}} J_z^+\cdot dl \right| \text{ if } J_z^+ > J_z^- \text{ near PIL} \\
\left |\oint_{B_z^{+,-}} J_z^-\cdot dl \right| \text{ if } J_z^- > J_z^+ \text{ near PIL}
\end{matrix}\right.
\label{DC}
\end{equation}

\textbf{
\begin{equation}
|RC|^{+,-} = \left\{\begin{matrix}
\left |\oint_{B_z^{+,-}} J_z^-\cdot dl \right| \text{ if } J_z^+ > J_z^- \text{ near PIL} \\
\left |\oint_{B_z^{+,-}} J_z^+\cdot dl \right| \text{ if } J_z^- > J_z^+ \text{ near PIL}
\end{matrix}\right.
\label{RC}
\end{equation}}

To determine the degree of current neutralization in the entire AR, \(|DC/RC|^{tot}\), we calculate the direct and return currents in each polarity separately, and then determine their cumulative ratio as illustrated in Equation \ref{DCRCtot},

\begin{equation}
|DC/RC|^{tot} = \frac{|DC|^+ + |DC|^-}{|RC|^+ + |RC|^-}
\label{DCRCtot}
\end{equation}

where the superscripts $+$ and $-$ denote values in positive and negative magnetic polarities.

Although the SHARP pipeline automatically selects regions of HMI magnetograms which contain ARs, for ARs that are not well-isolated or whose magnetic configuration is more complex, further steps need to be taken to isolate the magnetic flux necessary for determining our parameters of interest. We are primarily interested in magnetic flux which contributes to eruptions. For ARs which are more complex/less well-isolated, we locate the flaring part of the AR in extreme ultraviolet data from the Atmospheric Imaging Assembly \citep[AIA;][]{2012SoPh..275...17L} on \textit{SDO} and manually select the relevant sub-region that is involved in the eruption. Other model-dependent methods have also been developed to achieve this goal \citep[e.g.,][]{2017ApJ...846L...6L}. We discuss the uncertainties attributed to our region selection method in Section \ref{discussion}. 

\subsection{Calculation of PIL Magnetic Shear}\label{shear_calc}
To calculate the magnetic shear along the PIL of an AR, we first locate the PIL. We start by isolating all $B_z$ pixels whose magnitude is greater than 150 Gauss. We then smooth the masks of the positive and negative flux using a top-hat kernel and determine where the masks covering each polarity overlap. We then smooth the overlap mask with a Gaussian kernel to ensure we encompass the PIL. We use this mask to calculate PIL shear. An example of this mask is shown in the left panel of Figure \ref{shear_fig}. PIL shear is determined using methods described in previous studies \citep[e.g.,][]{1984SoPh...91..115H,1994ApJ...424..436W,2017ApJ...846L...6L}, which define PIL shear as the angle between the horizontal components of the observed magnetic field and a modeled potential magnetic field based on photospheric $B_z$. We base our code for computing PIL shear off of code written for the SHARP pipeline \citep{2014SoPh..289.3549B}. An example map of PIL shear values is shown in the right panel of Figure~\ref{shear_fig}.

\subsection{Calculation of Uncertainty}\label{var_calc}
We derive the statistical uncertainty in \(|DC/RC|\) and PIL shear from the statistical uncertainty in cylindrical equal area (CEA) coordinates (see \citet{2013arXiv1309.2392S} for a thorough description of the SHARP coordinate system and pipeline process). The HMI pipeline provides uncertainties as variances and covariances on the magnetic field magnitude, field inclination, field azimuth from the Stokes inversion process \citep{2014SoPh..289.3483H}. We propagate these errors through our measurements as follows.

We assume the uncertainties in an individual HMI native pixel follow a Gaussian distribution and first generate a correlated random sample ($N\sim100$) using the provided variances and covariances. This step creates an ensemble of new magnetograms. We then transform individual magnetograms in this ensemble into CEA coordinates, calculate the sample parameters of interest, and calculate the variance in the series of sample parameters. The CEA-coordinate transformation code is adapted from \textit{SolarSoft} IDL code (author X. Sun) and is available on Github\footnote{https://github.com/eavallon/CEA\_map}. The median statistical uncertainty for \(|DC/RC|^+\), \(|DC/RC|^-\), \(|DC/RC|^{tot}\), and PIL Shear is \(10^{-4}\), \(10^{-4}\), \(6.4\times10^{-5}\), and \(5.3\times10^{-3}\)$^{\circ}$ respectively. These uncertainties are \(99.5\%\) smaller than the systematic uncertainties presented in Table \ref{tab:big_tbl} and do not affect our results significantly.

The systematic uncertainty is computed using other methods. The primary source of systematic uncertainty is the periodicity introduced by the orbital velocity of \textit{SDO} around Earth \citep[for details, see][]{2014SoPh..289.3483H}. To quantify this uncertainty, we calculate the variance in our measurements within a 24-hour window. This is the uncertainty reported in Table \ref{tab:big_tbl}. We discuss further sources of systematic uncertainty in Section~\ref{discussion}.

\begin{table*}[t!]
    \centering
    \begin{tabular}{ccccccccc}
    \hline
    NOAA \# & CME & $|DC/RC|^+$ & $|DC/RC|^-$ & $|DC/RC|^{tot}$ & \shortstack{PIL Shear \\ ($^{\circ}$)} & \shortstack{Unsigned Flux \\ ($10^{22}$ Mx)}\footnote{Total magnetic flux on the day of flare/CME occurrence for flare-active ARs or on the day of 80\% maximum flux emergence for flare-quiet ARs calculated from \textit{SDO}/HMI SHARP magnetograms. This is calculated by taking the absolute value of \(B_z\) and summing all magnetic field above 200 Gauss in \(B_z\) maps to remove any contributions from noise. The magnetic field measurement is then converted to magnetic flux by multiplying it by the observed solar area in a single pixel.} & \shortstack{Flare \\ Index} & \shortstack{Sunspot \\ Classification}\footnote{Mt. Wilson sunspot classification on the day of flare/CME occurrence for flare-active ARs or on the day of 80\% maximum flux emergence for flare-quiet ARs from NOAA/USAF and \citet{2017ApJ...834...56T}. $\beta$ sunspot groups are simple bipoles with an easily identifiable PIL. $\gamma$ sunspot groups are more complex, with mixed positive and negative polarities and no easily identifiable PIL. $\delta$ sunspot groups contain at least one sunspot with multiple polarities inside the same sunspot penumbra.}  \\ \hline
	
	AR 11158 & Y & $1.18\pm0.02$ & $1.00\pm0.02$ & $1.19\pm0.02$ & $63.77\pm0.83$ & $3.24\pm0.01$ & 148 & $\beta\gamma$ \\
	AR 11261 & Y & $1.31\pm0.01$ & $1.60\pm0.01$ & $1.43\pm0.01$ & $49.97\pm3.01$ & $2.88\pm0.02$ & 27 & $\beta\gamma\delta$ \\
	AR 11429 & Y & $1.67\pm0.01$ & $1.62\pm0.01$ & $2.01\pm0.03$ & $53.89\pm0.98$ & $5.61\pm0.03$ & 442 & $\beta\gamma\delta$ \\
	AR 11515 & Y & $1.16\pm0.00$ & $1.26\pm0.00$ & $1.30\pm0.00$ & $58.68\pm0.80$ & $3.42\pm0.01$ & 217 & $\beta\gamma$ \\
	AR 11520 & Y & $1.32\pm0.00$ & $1.06\pm0.00$ & $1.17\pm0.00$ & $59.12\pm0.31$ & $10.48\pm0.05$ & 163 & $\beta\gamma\delta$ \\
	AR 11719 & Y & $1.06\pm0.00$ & $1.26\pm0.00$ & $1.33\pm0.01$ & $76.04\pm0.61$ & $2.10\pm0.01$ & 13 & $\beta\gamma$ \\
	AR 11890 & Y & $1.09\pm0.00$ & $1.13\pm0.00$ & $1.11\pm0.00$ & $47.60\pm4.08$ & $6.65\pm0.04$ & 370 & $\beta\gamma\delta$ \\
	AR 12242 & Y & $1.15\pm0.00$ & $1.13\pm0.00$ & $1.36\pm0.00$ & $49.98\pm1.03$ & $17.93\pm0.03$ & 155 & $\beta\gamma\delta$ \\
	AR 12297 & Y & $1.70\pm0.00$ & $1.72\pm0.00$ & $2.15\pm0.00$ & $62.78\pm2.91$ & $2.94\pm0.02$ & 337 & $\beta\gamma\delta$ \\
	AR 12371 & Y & $1.38\pm0.00$ & $1.32\pm0.00$ & $1.35\pm0.00$ & $66.62\pm0.63$ & $3.96\pm0.01$ & 91 & $\beta\gamma\delta$ \\
	AR 12673 & Y & $1.62\pm0.01$ & $1.39\pm0.01$ & $1.48\pm0.01$ & $52.71\pm3.34$ & $5.51\pm0.03$ & 399 & $\beta\gamma\delta$ \\ 
	AR 11166 & N & $1.17\pm0.00$ & $1.06\pm0.00$ & $1.11\pm0.00$ & $61.04\pm0.52$ & $3.35\pm0.01$ & 136 & $\beta\gamma\delta$ \\
	AR 11302 & N & $1.06\pm0.00$ & $1.16\pm0.00$ & $1.10\pm0.00$ & $48.93\pm3.21$ & $5.55\pm0.03$ & 97 & $\beta\gamma\delta$ \\
	AR 11339 & N & $1.07\pm0.00$ & $1.05\pm0.00$ & $1.06\pm0.00$ & $43.79\pm2.18$ & $8.60\pm0.06$ & 83 & $\beta\gamma\delta$ \\
	AR 12192 & N & $1.02\pm0.00$ & $1.09\pm0.00$ & $1.05\pm0.00$ & $60.00\pm2.38$ & $16.64\pm0.08$ & 594 & $\beta\gamma\delta$ \\ \hline

	AR 11776 & N & $0.98\pm0.00$ & $1.11\pm0.01$ & $1.05\pm0.00$ & $44.64\pm3.43$ & $1.21\pm0.00$ & 4 & $\beta\delta$ \\
	AR 11784 & N & $0.98\pm0.00$ & $0.95\pm0.01$ & $0.97\pm0.00$ & $50.17\pm4.39$ & $1.24\pm0.00$ & 1 & $\beta\gamma\delta$ \\
	AR 11887 & N & $1.10\pm0.00$ & $1.12\pm0.00$ & $1.10\pm0.00$ & $55.15\pm1.43$ & $1.16\pm0.00$ & 0 & $\beta\gamma$ \\
	AR 11957 & N & $1.05\pm0.01$ & $1.26\pm0.01$ & $1.15\pm0.01$ & $45.26\pm14.21$ & $1.75\pm0.00$ & 0 & $\beta\gamma$ \\
	AR 12071 & N & $0.99\pm0.00$ & $1.16\pm0.00$ & $1.05\pm0.00$ & $53.42\pm1.77$ & $3.49\pm0.00$ & 3 & $\beta\gamma$ \\
	AR 12082 & N & $1.05\pm0.00$ & $1.08\pm0.00$ & $1.05\pm0.00$ & $46.92\pm8.63$ & $3.07\pm0.00$ & 1 & $\beta$ \\
	AR 12100 & N & $1.15\pm0.00$ & $1.04\pm0.00$ & $1.09\pm0.00$ & $51.80\pm5.17$ & $1.27\pm0.00$ & 2 & $\beta$ \\
	AR 12121 & N & $1.04\pm0.00$ & $1.08\pm0.00$ & $1.06\pm0.00$ & $52.86\pm4.67$ & $2.80\pm0.00$ & 1 & $\beta$ \\
	AR 12203 & N & $0.94\pm0.02$ & $1.33\pm0.05$ & $1.14\pm0.03$ & $80.98\pm1.03$ & $0.87\pm0.00$ & 0 & $\beta$ \\
	AR 12239 & N & $1.14\pm0.00$ & $1.01\pm0.00$ & $1.07\pm0.00$ & $46.67\pm9.13$ & $0.90\pm0.00$ & 0 & $\beta$ \\
	AR 12244 & N & $1.35\pm0.02$ & $1.27\pm0.01$ & $1.31\pm0.01$ & $56.56\pm2.35$ & $0.94\pm0.00$ & 0 & $\beta$ \\
	AR 12545 & N & $1.60\pm0.05$ & $1.06\pm0.01$ & $1.28\pm0.01$ & $32.97\pm13.65$ & $0.82\pm0.00$ & 0 & $\beta$ \\
	AR 12629 & N & $1.00\pm0.00$ & $1.06\pm0.00$ & $1.03\pm0.00$ & $43.98\pm2.19$ & $0.87\pm0.00$ & 0 & $\beta$ \\
	AR 12683 & N & $1.02\pm0.00$ & $1.02\pm0.00$ & $1.02\pm0.00$ & $46.96\pm3.20$ & $2.59\pm0.00$ & 0 & $\beta$ \\
	AR 12715 & N & $1.27\pm0.01$ & $1.19\pm0.01$ & $1.23\pm0.01$ & $38.50\pm7.71$ & $0.73\pm0.00$ & 0 & $\beta$ \\
    \end{tabular}
    \caption{CME presence, $|DC/RC|$ for each magnetic polarity and the entire AR, PIL shear, Unsigned Flux, Flare Index, and Sunspot Classification for the full sample of 30 ARs. The top ARs are flare-active and the bottom are flare-quiet. The Table is sorted by CME-activity and NOAA AR number. We report the systematic uncertainty from the orbit of \textit{SDO} around Earth as the associated error for each parameter. The statistical uncertainty for each parameter is \(0.01\%\) of the systematic uncertainty and does not contribute significantly to the overall uncertainty in our results.}
    \label{tab:big_tbl}
\end{table*}

\section{Results}\label{results}
We summarize the results for our full sample of flare-active and flare-quiet ARs in Table \ref{tab:big_tbl}. Our sample is grouped by flare activity and organized based on their CME activity and their National Oceanic and Atmospheric Administration (NOAA) AR number. Our presented results include the presence of CMEs, \(|DC/RC|\) for each magnetic polarity and the entire AR, PIL shear, total unsigned flux, flare index, and sunspot classification. For flare-active ARs, the values are reported on the day of CME onset or the day of the maximum magnitude flare produced by that AR. For flare-quiet ARs, the values are reported on the day when the flux is at 80\% its maximum value. We refer to these values when we present a single number for \(|DC/RC|\), PIL shear, or unsigned flux.

\begin{figure*}
    \centering
    \includegraphics[width=0.9\textwidth]{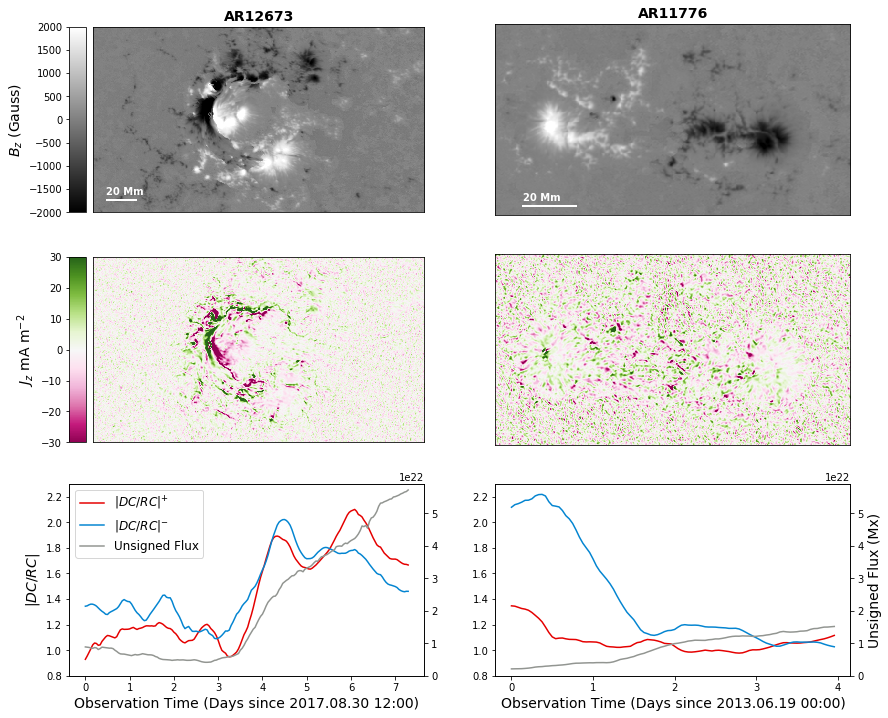}
    \caption{Flare-active AR 12673 (left) and flare-quiet AR 11776 (right). The top panels show the vector magnetogram for the \(z\) (vertical) component of the magnetic field. The middle panels show the derived map of \(J_{z}\), the vertical component of the current density. The bottom panels show \(|DC/RC|\) in the positive (red) and negative (blue) polarities of each AR and the unsigned flux (gray) over the on-disk lifetime of each AR. The \(|DC/RC|\) plots are smoothed by a boxcar kernel of 12 hours to more clearly show trends. Error bars are too small to be visible on the plot and are thus omitted. Error values can be found in Table \ref{tab:big_tbl}.}
    \label{fig:pretty_ars}
\end{figure*}

An example of a flare-active and a flare-quiet AR is shown in Figure \ref{fig:pretty_ars}, which shows the magnetogram (top), derived \(J_z\) map (middle), and \(|DC/RC|\) values for each magnetic polarity and the unsigned flux over the entire on-disk lifetime of each AR (bottom). The highlighted ARs are AR 12673 (flare-active) and AR 11776 (flare-quiet). The ARs in our flare-quiet sample show similar trends to AR 11776 in their on-disk evolution of \(|DC/RC|^+\) and \(|DC/RC|^-\). The bottom right panel of Figure \ref{fig:pretty_ars} shows that the electric currents become more neutralized as magnetic flux emerges (i.e., as unsigned flux increases in magnitude). We see the same trend in 60\% of ARs in the flare-quiet sample. A similar trend in \(|DC/RC|\) is not observed in our flare-active sample. We also do not observe any other trends in \(|DC/RC|\) in the flare-active sample. 

We do, however, observe similarities in the structure of \(J_z\) in our flare-active sample. The middle left panel of Figure \ref{fig:pretty_ars} shows the \(J_z\) map for AR 12673. We can see that the positive and negative currents have a coherent structure around the PIL of this AR, indicating the presence of ``current ribbons'' as in a coherent flux rope. We observe similar current ribbons in 87\% of flare-active ARs and in 6\% of flare-quiet ARs. The ARs in our sample that do not exhibit current ribbons have similar disordered current distributions to AR 11776, whose \(J_z\) map is shown in the middle right panel of Figure \ref{fig:pretty_ars}. These ARs have current distributions which do not have any noticeable structure, and the positive and negative currents appear to be evenly distributed.  

\begin{figure*}[ht!]
    \centering
    \includegraphics[width=0.9\textwidth]{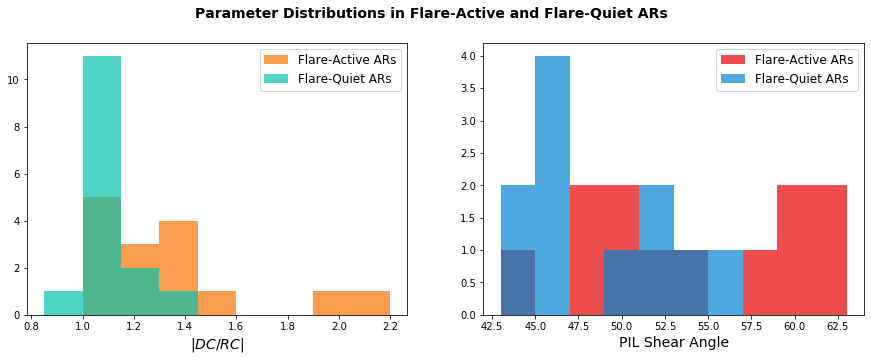}
    \caption{Histograms showing the distribution of \(|DC/RC|^{tot}\) (left) and PIL shear (right) values in flare-active (orange/red) and flare-quiet (green/blue) ARs. The plotted values are the same as those presented in Table \ref{tab:big_tbl}. Both \(|DC/RC|^{tot}\) and PIL shear differ between flare-active and flare-quiet ARs.}
    \label{fig:hist}
\end{figure*}

We arbitrarily define current-neutralized ARs as ARs with a \(|DC/RC|^{tot}\) value less than 1.10. Based on our full sample, we find that flare/CME-active ARs tend to be less current-neutralized than flare/CME-quiet ARs. 73\% of flare-quiet ARs are current-neutralized, while only 13\% of flare-active ARs are current neutralized. The distribution of \(|DC/RC|^{tot}\) for our entire sample is shown in the left panel of Figure \ref{fig:hist}. A Kolmogorov-Smirnov (KS) test on these distributions yields a $p$-value of 0.051, which indicates that the \(|DC/RC|^{tot}\) values for flare-active and flare-quiet ARs are sampled from different distributions at a 90\% significance level but are not sampled from different distributions at a 95\% significance level. 

The values for PIL shear also differ between the flare-active and flare-quiet sample. The distribution of PIL shear for our entire sample is shown in the right panel of Figure \ref{fig:hist}. A KS test on these distributions yields a $p$-value of 0.075, indicating that PIL shear values for flare-active and flare-quiet ARs are sampled from different distributions at a 90\% significance level but are not sampled from different distributions at a 95\% significance level.

\begin{figure}
    \centering
    \includegraphics[width=0.46\textwidth]{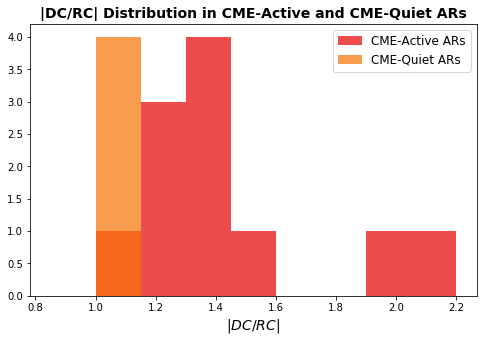}
    \caption{Histogram showing the distribution in $|DC/RC|^{tot}$ values for CME-active (red) and CME-quiet (orange) ARs in our flare-active sample. CME-active ARs are less current-neutralized than CME-quiet ARs.}
    \label{fig:cme_hist}
\end{figure}

We also report on the CME-productivity, flare index, and sunspot classification of the ARs in our sample in Table \ref{tab:big_tbl}. Flare-active ARs produce more CMEs, with 73\% of our flare-active sample producing at least one CME. Flare-active ARs which do not produce any CMEs also tend to be more current-neutralized than those that do produce CMEs. A histogram of \(|DC/RC|^{tot}\) for CME-active and CME-quiet flare-active ARs is shown in Figure \ref{fig:cme_hist}. Such flare-active and CME-quiet ARs have also been more thoroughly characterized in previous studies \citep[see e.g.,][]{2015ApJ...804L..28S}. All flare-active and most CME-active ARs also have higher flare indices than flare/CME-quiet ARs, indicating that they produce more flares of a higher magnitude. Flare-active ARs finally have more complex magnetic configurations than flare-quiet ARs, as indicated by the Sunspot Classification column in Table \ref{tab:big_tbl}. All flare-active ARs are classified as $\gamma$ sunspot groups, with 80\% also exhibiting a $\delta$ spot. This contrasts with the flare-quiet sample, where only 33\% of these ARs have a sunspot classification of $\gamma$ or $\delta$. 

\begin{figure*}[ht!]
    \centering
    \includegraphics[width=0.6\textwidth]{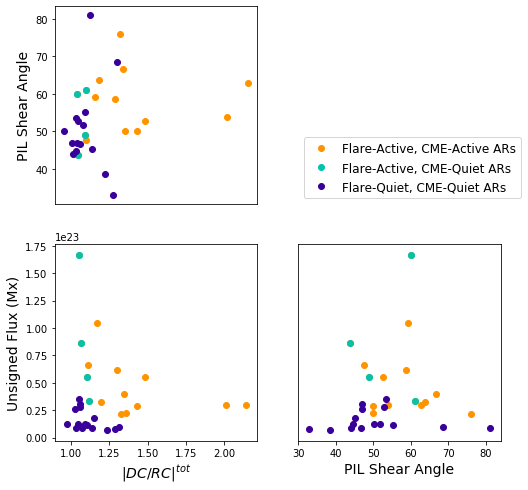}
    \caption{Plot showing comparisons between \(|DC/RC|^{tot}\), PIL shear, and unsigned flux for flare-active/CME-active (orange), flare-active/CME-quiet (green), and flare-quiet/CME-quiet (purple) ARs. There is no significant linear correlation between any of these parameters based on Pearson correlation coefficients of -0.071 for unsigned flux and \(|DC/RC|^{tot}\), 0.221 for PIL shear and \(|DC/RC|^{tot}\), and 0.105 for PIL shear and unsigned flux. Error bars smaller than the circles are omitted, but can be found in Table \ref{tab:big_tbl}.}
    \label{fig:corner}
\end{figure*}

We also compare \(|DC/RC|^{tot}\) and PIL shear with the total unsigned flux in each AR to determine whether any correlations are present. Scatter plots showing these comparisons are shown in Figure \ref{fig:corner}. We can see in the bottom left and bottom right panels in Figure \ref{fig:corner} that the unsigned flux is higher for flare-active ARs than for flare-quiet ARs, further indicated by a KS test $p$-value of 0.0002. The distribution in unsigned flux also follows the distribution in \(|DC/RC|^{tot}\), but does not linearly correlate with it, based on a Pearson correlation coefficient of -0.071. Although \(|DC/RC|^{tot}\) and PIL shear differ significantly between flare-active and flare-quiet ARs, we see no clear correlation between these values, as shown in the top left panel of Figure \ref{fig:corner} and by a Pearson correlation coefficient of 0.221. We also see no clear correlation between PIL shear and unsigned flux, as shown in the bottom right panel of Figure \ref{fig:corner} and by a Pearson correlation coefficient of 0.105. 

\section{Discussion \& Conclusions}\label{discussion}
Based on our analysis of 15 flare-active and 15 flare-quiet ARs, we find that flare-active/CME-active ARs are less current-neutralized than flare-active/CME-quiet ARs or flare-quiet/CME-quiet ARs. These results support the findings in \citet{2017ApJ...846L...6L}. The difference in \(|DC/RC|\) for flare-active/CME-active ARs and flare-active/CME-quiet ARs indicates that the degree to which AR currents are neutralized are systematically different in these two populations. Prediction of flare/CME production based on \(|DC/RC|\) alone, however, can be difficult as there is some overlap between the two populations.

We note that flares and CMEs do not always occur together even for the most energetic events. This is especially apparent for AR 12192, which has an exceptionally high flare index of 594, but does not produce any CMEs \citep{2015ApJ...804L..28S}.

We also find that CME-eruptive ARs exhibit defined current ribbons while CME-quiet ARs do not, as shown in Figure \ref{fig:pretty_ars}, suggesting the presence of a pre-eruption coherent flux rope. This is a necessary condition for CME eruption in models that evoke the kink or torus instability \citep{forbes_2010}. This also follows the distribution of currents shown in Figure \ref{flux_rope}, with the direct currents confined to the center of the flux rope and return currents in a surrounding sheath. 

The presence of non-neutralized currents in our sample of ARs is additionally consistent with previous simulations which argue that return currents get trapped beneath the photosphere during flux emergence, thus leading to the presence of a non-neutralized current along with substantial PIL shear.

\begin{figure*}[t!]
    \centering
    \includegraphics[width=0.9\textwidth]{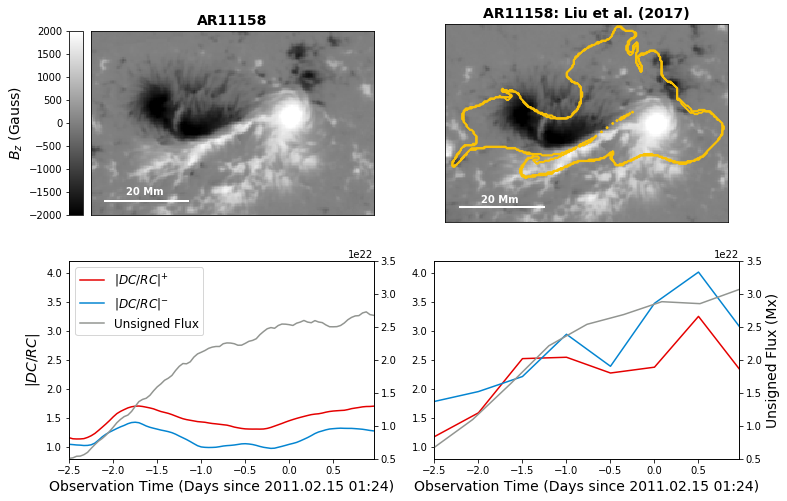}
    \caption{A comparison between our region selection method (left) and the region selection method used in \citet[][right]{2017ApJ...846L...6L} for AR 11158. The top panels show magnetograms of \(B_z\) and the bottom panels show plots of \(|DC/RC|\) over the on-disk lifetime of AR 11158. The mask used by \citet{2017ApJ...846L...6L} is indicated by the yellow line in the top right panel. We perform our measurements over the entire region plotted in the top left panel. The computed value of \(|DC/RC|\) is dependent on the region selection method used.}
    \label{ar_mask}
\end{figure*}

\begin{figure*}
    \centering
    \includegraphics[width=0.9\textwidth]{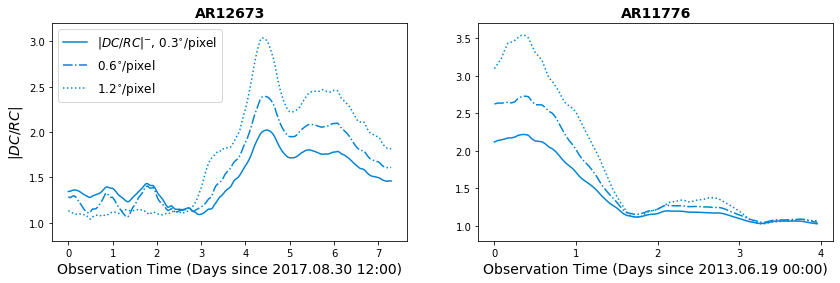}
    \caption{The effect of magnetogram resolution on our calculated value of \(|DC/RC|^-\) for flare-active AR 12673 and flare-quiet AR 11776. We modify the resolution by binning the original magnetogram of each AR. We find that \(|DC/RC|^-\) increases with decreasing resolution. We note that although AR 11776 reaches values similar to non-neutralized ARs, those values are reached prior to any significant flux emergence (see bottom right panel of Figure \ref{fig:pretty_ars}) and are likely dominated by noise.}
    \label{binning}
\end{figure*}

The decreasing trends observed in the early evolution of \(|DC/RC|\) in flare/CME-quiet ARs, to our knowledge, have not been addressed in previous studies. Previous simulations show that non-neutralized currents develop alongside flux emergence \citep[e.g.,][]{1991ApJ...381..306M,1995ApJ...451..391M,2000ApJ...545.1089L}, while the ARs in our flare-quiet sample become more neutralized as magnetic flux emerges. This behavior can also be seen in AR 11072, which is discussed in \citet{2017ApJ...846L...6L}. There is some indication of this behavior in \citet{2014ApJ...782L..10T} and \citet{2018ApJ...864...89K}. However, none of these simulations indicate that currents should become neutralized alongside flux emergence. One possible explanation is that the measurement of substantial currents prior to significant flux emergence is due to extraneous flux patches, or is further affected by noise. Improved uncertainty constraints or existing simulations of CME-quiet ARs should clarify this issue. 

We find a significant difference in PIL shear for flare/CME-active and flare/CME-quiet ARs as well. This supports the findings in \citet{2017ApJ...846L...6L} and previous simulations \citep{2015ApJ...810...17D}. Our results, however, contradict other theoretical studies which show that the presence of non-neutralized currents in ARs is directly correlated with PIL shear \citep[e.g.,][]{2004SoPh..219...87W}. A larger sample could clarify this discrepancy.

Our sample was selected based on the presence of flares in ARs. Since our results primarily have implications for the CME-productivity of ARs, selecting another large sample of low M-class to C-class flaring ARs with and without CMEs could provide further conclusions. Data-driven models of CME-active and CME-quiet ARs can also be used to further characterize these ARs \citep{2015SpWea..13..369F}. High resolution data based on multi-line observations from the upcoming \textit{Daniel K. Inouye Solar Telescope} (\textit{DKIST}) may allow for measurement of current distribution at different heights in the solar atmosphere \citep{2016AN....337.1064T}. This should further test the trapping scenario described in \citet{2014ApJ...782L..10T}.

Along with variations introduced from the orbit of \textit{SDO} around Earth mentioned in Section \ref{var_calc}, our selection method for our integration region is also a source of uncertainty (see Section \ref{dcrc_calc}). To quantify how much this affects our measurements, we included in our sample one of the ARs in \citet{2017ApJ...846L...6L}, AR 11158. \citet{2017ApJ...846L...6L} used a nonlinear force-free extrapolation to determine a magnetically closed central region around the flaring PIL. A comparison between this model-selected region and our selected region and the effects on our measurement of \(|DC/RC|\) is shown in Figure \ref{ar_mask}. We find that our values for \(|DC/RC|\) are a factor of two smaller than those reported in \citet{2017ApJ...846L...6L}. We expect the alternative method will increase the \(|DC/RC|\) values in flare/CME-active ARs, while affecting less the quiet ones. Thus it will likely strengthen our conclusion.

We additionally consider image resolution effects as a source of uncertainty. To analyze the effects of image resolution on our measurement of \(|DC/RC|^-\), we decrease the resolution of the magnetograms of two example ARs, AR 12673 and AR 11776 by a factor of two and a factor of four, the results of which are shown in Figure \ref{binning}. We find that decreasing the image resolution increases the values of \(|DC/RC|^-\), \(|DC/RC|^+\), and \(|DC/RC|^{tot}\) in general, once a significant amount of flux has emerged (after day 3 for AR 12673 and day 1.5 for AR 11776). The spatial averaging smooths out small scale field structure, which creates a larger impact on the minority/less structured return currents. This is consistent with the findings from \citet{2012ApJ...761...61G}. The authors found that decreasing the resolution increased the current "non-neutrality factor", an index similar to \(|DC/RC|\). \(|DC/RC|\) behaves in a less predictive way prior to significant flux emergence. It is unclear why that is, but the result is indeed more susceptible to noise and systematic effects. We caution against using \(|DC/RC|\) values at the early emergence phase.

This sample, while a larger sample than what was used in previous studies, is still subject to errors introduced in statistical analyses due to the small sample size. Future studies with larger samples can help to further understand the parameters measured in this study and their distribution within the entire population of observed ARs.

\acknowledgements
We thank Monica Bobra, Manolis Georgoulis, Kalman Knizhnik, James Leake, Yang Liu, and Tibor T\"{o}r\"{o}k for their informative and helpful discussions. The \textit{SDO} data are courtesy of NASA, the \textit{SDO}/HMI, and AIA science team. This work was supported by NSF award \#1848250.

\end{CJK*}

\bibliographystyle{aasjournal}
\bibliography{final}

\end{document}